# Gate-versus defect-induced voltage drop and negative differential resistance in vertical graphene heterostructures

Tae Hyung Kim†, Juho Lee†, Ryong-Gyu Lee, and Yong-Hoon Kim*

*School of Electrical Engineering, Korea Advanced Institute of Science and Technology (KAIST), 291 Daehak-ro, Yuseong-gu, Daejeon 34141, Korea.*

E-mail: y.h.kim@kaist.ac.kr

**Abstract**

Vertically stacked two-dimensional (2D) van der Waals (vdW) heterostructures based on graphene electrodes represent a promising architecture for next-generation electronic devices. However, their first-principles characterizations have been so far mostly limited to the equilibrium state due to the limitation of the standard non-equilibrium Green's function approach. To overcome these challenges, we introduce a non-equilibrium first-principles calculation method based on the recently developed multi-space constrained-search density functional formalism and apply it to graphene/few-layer hexagonal boron nitride (hBN)/graphene field-effect transistors. Our explicit finite-voltage first-principles calculations show that the previously reported negative differential resistance (NDR) current-bias voltage characteristics can be produced not only from the gating-induced mismatch between two graphene Dirac cones but from the bias-dependent energetic shift of defect levels. Specifically, for a carbon atom substituted for a nitrogen atom ($C_N$) within inner hBN layers, the increase of bias voltage is found to induce a self-consistent electron filling of in-gap $C_N$ states, which leads to changes in voltage drop profiles and symmetric NDR characteristics. On the other hand, with a $C_N$ placed on outer interfacial hBN layers, we find that due to the pinning of $C_N$ levels to nearby graphene states voltage drop profiles become bias-independent and NDR peaks disappear. Revealing hitherto undiscussed non-equilibrium behaviors of atomic defect states and their critical impact on device characteristics, our work points towards future directions for the computational design of 2D vdW devices

## 1. INTRODUCTION

The recent development of van der Waals (vdW) heterostructures based on diverse combinations of two-dimensional (2D) crystals opened up exciting possibilities to explore novel physical phenomena as well as device applications[1,2]. Here, much effort has been devoted to identify device architectures that maximally utilize novel properties emerging from hybrid forms of 2D vdW materials in nonequilibrium conditions[3-5]. For example, the junction configuration with few-layer 2D semiconductors or insulators sandwiched between graphene electrodes represents a promising platform from which phenomena such as the chiral quantum state[6,7], giant tunneling magnetoresistance[8,9], and negative differential resistance (NDR)[7,10-12] can be explored.

For the continued development of novel physics and advanced devices based on atomically thin 2D vdW heterostructures, it will be necessary to properly understand how atomic-scale details affect the heterojunction electronic structures and quantum transport properties under finite-bias non-equilibrium conditions. However, the standard first-principle approach combining density functional theory (DFT) and non-equilibrium Green's function (NEGF) formalism within the Landauer framework[13] has intrinsic shortcomings in simulating 2D vdW heterojunctions under finite-bias conditions (Fig. 1a). Specifically, taking as an example the junction configuration in which a few-layer hexagonal boron nitride (hBN) channel is sandwiched by graphene electrodes, DFT-NEGF formalism requiring electrodes to be repeated semi-infinitely along the transport direction forces one to replace single-layer graphene electrodes with infinite-layer graphite counterparts (Fig. 1b) or to adopt an hBN nanoribbon channel model in combination with edged graphene electrodes (Fig. 1c). Both approaches then suffer from important shortcomings in that graphene and graphite electrodes will behave differently in the former case[12,14-16] and that it will be difficult to avoid edge-shape and width dependences of hBN nanoribbons in the latter case[17,18]. Accordingly, most theoretical studies currently resort to semi-classical approaches such as the Bardeen transfer Hamiltonian formalism[19,20] to handle the laterally periodic 2D device geometry (Fig. 1d), eliminating the possibility of interpreting and predicting effects that involve atomistic details in an *ab initio* manner[10,21-26].

In this work, extending the recently developed multi-space constrained-search DFT (MS-DFT) formalism[27-29], we analyze the non-equilibrium electronic structure and quantum transport properties of vertical graphene/few-layer hBN/graphene field effect transistors (FETs) at the atomic level. The MS-DFT formalism is established by invoking a microcanonical viewpoint or employing finite-sized metallic electrodes, in contrast to the semi-infinite metallic electrodes from the grand canonical picture of DFT-NEGF. We first explain how the MS-DFT formalism can be straightforwardly extended such that for 2D vdW heterojunctions incorporating graphene and other 2D electrodes quantum transport properties including gating effects can be treated within the true vertical 2D FET geometry (Fig. 1d left panel). Next, applying MS-DFT to graphene/hBN/graphene FETs, we analyze the gate-induced negative differential resistance (NDR) current density-bias voltage ($J - V_b$) characteristic in the pristine hBN channel case and quantify the hBN layer number-dependent quantum capacitance effect of graphene. We then demonstrate how a point defect introduced into the hBN channel critically affects finite-bias junction electronic structure and device operations. Specifically, we consider a carbon atom substituted for a nitrogen atom ($C_N$) in an inner hBN layer (Fig. 1d right panel)[30], and show that it can lead to an NDR signal even in the absence of gating effects. Its origins will be explained in terms of the self-consistent charging and upshifting of in-gap $C_N$ defect levels with the increase of bias voltage, which is accompanied by the change in the behavior of the spatial electrostatic potential or "voltage" drop profile. On the other, for the $C_N$ defect introduced into an outer interfacial hBN layer, we find that the NDR peak disappears and the current density decreases by an order of magnitude. We will show that these drastic changes result from the pinning of in-gap $C_N$ levels to the neighboring graphene states, which translates into a fixed spatial voltage drop behavior. These non-equilibrium atomic details that unravel the mechanism of hitherto unexplained symmetric NDR $J - V_b$ characteristics hint at new opportunities in the understanding and design of 2D vdW heterojunction devices.

## 2. RESULTS AND DISCUSSION

*2.1 First-principles approach for the simulation of 2D vdW FETs*

As an alternative to the standard DFT-NEGF approach, we previously established the MS-DFT formalism by adopting a microcanonical picture or finite electrodes and mapping the finite-bias electron quantum transport process to the drain-to-source optical excitation counterpart[27-29]. An important practical implication for this work is then that, unlike the DFT-NEGF scheme that is based on the Landauer picture and thus requires semi-infinite electrodes, MS-DFT provides a natural framework to perform first-principles calculations for vertical 2D vdW FETs based on graphene electrodes. We here explain how the MS-DFT approach can be extended such that (i) the gate voltage $V_g$ can be included in addition to the bias voltage $V_b$, and (ii) transmissions at finite $V_b$ and/or $V_g$ can be calculated without recovering the Landauer picture or introducing semi-infinite electrodes.

Regarding (i), we reiterate that within the MS-DFT approach the finite $V_b$ was embodied as a constraint $eV_b = \mu_L - \mu_R$ for the total-energy minimization. Similarly, after the left electrode L/channel C/right-electrode R/gate G partitioning (see also **Methods**), the gate bias $V_g$ can be straightforwardly implemented as an additional constraint of $-eV_g = \mu_g - (\mu_L + \mu_R)/2$, where $\mu_g$ stands for the electrochemical potential of G, which precedes the $V_b$ constraint. For the practical implementation of this scheme, as schematically shown in Fig. 1d, we adopted an additional Au monolayer as the bottom gate electrode and extended the spatial tracing and occupation constraining of Kohn-Sham states to the gate electrode region. A vacuum space of 10 Å was introduced between the graphene/hBN/graphene junction and the Au electrode to ensure that the gate leakage current is completely suppressed. This corresponds to the 3.9 nm equivalent oxide thickness (EOT) which is defined as

$$\text{EOT} = \left(\frac{\varepsilon_{SiO_2}}{\varepsilon_{gate}}\right) t_{gate}, \quad (1)$$

where $\varepsilon_{SiO_2}$ and $\varepsilon_{gate}$ are the dielectric constants of SiO$_2$ ($\varepsilon_{SiO_2} = 3.9\,\varepsilon_0$) and vacuum ($\varepsilon_{gate} = \varepsilon_0$), respectively, and $t_{gate}$ is the vacuum thickness (1 nm).

To achieve (ii), after obtaining a non-equilibrium electronic structure within MS-DFT calculations, we invoked as a post-processing step the matrix Green's function formalism[31,32] and calculated finite-bias transmission coefficients according to[13]

$$T(E; V_b, V_g) = Tr[\boldsymbol{a}_L \mathbf{M} \boldsymbol{a}_R \mathbf{M}^\dagger], \quad (2)$$

where $\boldsymbol{a}_{L(R)}$ is the spectral function in the L (R) contact and $\mathbf{M} = \mathbf{x}_L^\dagger \mathbf{G} \mathbf{x}_R$. In computing $\boldsymbol{a}_{L(R)}$, since electrode unit cells semi-infinitely repeated along the transport direction do not exist anymore for the present vertical 2D vdW heterojunction model (Fig. 1d), we replaced the surface Green's function $\mathbf{g}_s^{L(R)}$ with the region L (R) retarded Green's function $\mathbf{G}$ calculated from the junction model itself (orange boxes in Fig. 1d). Here, we introduce a constant broadening factor (~0.025 eV), which originally enters into the construction of $\mathbf{g}_s$ for the semi-infinite electrode case and physically represents the nature of electrons incoming from (outgoing into) the source (drain) electrode. It should be noted that, while the matrix element $\mathbf{M}$ approximately corresponds to the tunneling matrix in the Bardeen transfer Hamiltonian approach[19], it now properly accommodates the impact of coupling between the channel and electrodes and their atomistic details[13]. Once the transmission functions were obtained, the current density-bias voltage ($J$-$V_b$) characteristic under $V_g$ was calculated using the Landauer-Büttiker formula[13],

$$J(V_b, V_g) = \frac{2e}{h} \int_{\mu_L}^{\mu_R} T(E; V_b, V_g)[f(E - \mu_R) - f(E - \mu_L)] dE, \quad (3)$$

where $f(E - \mu) = 1/\{1 + exp((E - \mu)/k_B T)\}$ is the Fermi-Dirac distribution function.

We implemented these novel MS-DFT functionalities within the SIESTA package[33], which has been extensively employed for the DFT-NEGF program development[34-37]. For the benchmark, because the laterally infinite vertical 2D vdW FET models shown in Fig. 1d cannot be handled with existing DFT-NEGF codes, we instead considered the graphite/few-layer hBN/graphite heterojunction case shown in Fig. 1b that can be also handled within DFT-NEGF. In addition to checking the computational convergence trend, we confirmed the excellent agreement between MS-DFT and DFT-NEGF transmission data (see Supplementary Fig. 1).

## 2.2 Gate-induced asymmetric NDR J-$V_b$ curves and linear background currents

We first applied MS-DFT to 2D vdW heterojunctions composed of few-layer pristine hBN channels sandwiched between graphene electrodes, a prototype configuration for the experimental realization of vertical 2D vdW tunneling FETs and the main focus of this work[10,11,21], we then set $V_g = 0$ V and calculated the finite-bias electronic and quantum transport properties. Varying the number of hBN layers ($N_{BN}$) from $N_{BN} = 2$ to $N_{BN} = 5$, as shown in the left panel of Fig. 2a, we obtain the linear $J$-$V_b$ characteristics and an exponential current decrease with increasing $N_{BN}$ with a decay constant of 0.718 Å$^{-1}$. These results are overall in agreement with the mechanism of quantum tunneling in biased graphene/2D semiconductor or insulator/graphene vdW heterojunctions, which has been extensively discussed in previous reports[21-23,38]. Given the constraints of energy and momentum conservation, as schematically shown in the right panel of Fig. 2a, tunneling between two shifted graphene Dirac cones can occur for a single ring of $\vec{k}$ points. This ring is located at $E = (\mu_R + \mu_L)/2$ under $V_g = 0$ V, and the current will linearly increase as the circumference of the ring linearly expands with $V_b$.

The top panel of Fig. 2b shows for the trilayer (3L) hBN case the plane-averaged bias-induced modification of the Hartree electrostatic potential difference at $V_b = 1.0$ V, calculated according to

$$\Delta \bar{v}_H(\vec{r}) = \bar{v}_H(\vec{r}; V_b, V_g) - \bar{v}_H(\vec{r}; 0, 0), \quad (4)$$

The corresponding exchange-correlation potential variations are negligible and will not be explicitly discussed below (for details, see Supplementary Fig. 2). In nanoscale junctions, as will be discussed shortly, the behaviors of the quasi-Fermi level $\mu$ drop and the electrostatic potential $\Delta \bar{v}_H$ drop are in general different[28,39], and below we will address the latter as the "voltage drop".

For the LG/3L hBN/RG junction, the LG-RG $\Delta \bar{v}_H$ offset calculated at $V_b = 1.0$ V was 0.52 eV, which is smaller than the applied electrochemical potential difference $\mu_R - \mu_L = eV_b = 1.0$ eV. Specifically, the mismatch between the variation of the electrochemical potential and that of the electrostatic potential is a manifestation of the graphene quantum capacitance[42,43]. To quantify the hBN layer number-dependent quantum capacitance effects, we additionally measured the $V_b = 1.0$ V $\Delta \bar{v}_H$ values for the tetralayer (4L) and pentalayer (5L) cases and obtained 0.56 eV and 0.62 eV, respectively (for the 5L hBN case data, see Supplementary Fig. 3).

To further analyze the corresponding non-equilibrium junction electronic structure, we visualize the finite-$V_b$ band structures of the LG/3L hBN/RG junction projected onto the LG and RG and show them in Fig. 2c. Note that the possibility to straightforwardly analyze finite-$V_b$ electronic structures using the standard DFT analysis methods is a unique strength of the MS-DFT formalism. Then, in addition to the 0.52 eV offset of the LG and RG Dirac cones, we identify as a notable feature that the bands of several adjacent hBN layers (marked by purple down triangles) are projected onto the LG (RG) band.

Next, forming the LG/3L hBN/RG FET model by introducing a bottom gate electrode (gate electrode placed on the LG side), we studied the effects of gating on graphene-based vertical 2D vdW FETs. In the $J - V_b$ characteristics calculated for $V_g = +2.5$ eV, $V_g = -0.5$ eV, and $V_g = -2.5$ eV shown in Fig. 2d, we find in good agreement with previous experimental data[10] pronounced asymmetric NDR signals superimposed over linear background currents. Partial penetration of an external field is another general characteristic of the quantum capacitance of 2D electron gas,[44] which results in nontrivial voltage drop profiles. The $\Delta \bar{v}_H$ curves obtained under $V_g = -2.5$ V (Fig. 2e top panels) show that a voltage drop between LG and RG of $\Delta \bar{v}_H = +0.11$ eV already appears at $V_b = 0$ V (see Supplementary Fig. 4) and it decreases (increases) to $-0.42$ eV ($+0.84$ eV) at $V_b = -1.0$ V ($+1.0$ V). This asymmetric $J - V_b$ characteristics can be understood by observing the corresponding $\Delta \bar{\rho}$ plots (Fig. 2e bottom panels): Unlike in the $V_b = +1.0$ V case, the application of $V_b = -1.0$ V with the same polarity as $V_g = -2.5$ V results in the depletion of LG electrons in screening the gate electric field and accordingly a reduced capacity to support the electrochemical potential difference of $V_b = -1.0$ V.

As shown in Fig. 2f, projecting $V_g = -2.5$ V band structures to LG and RG at varying $V_b$ additionally reveals the detailed operation mechanisms of vertical graphene-based vdW tunneling FETs that exhibit NDR features (for the $V_g = +2.5$ V 3L hBN case and the $V_g = -2.5$ V 4L hBN case, see Supplementary Figs. 5-6, respectively). As is evident in the $V_b = 0$ V case (Fig. 2f, fourth left panel), the application of $V_g = -2.5$ V induces the hole doping of both LG and RG and, given the bottom gate device geometry, the hole doping level is higher for LG and the misalignment of the LG and RG Dirac cones results in marginal transmission spectra (Fig. 2f, fourth right panel). Then, when $V_b$ increases positively (negatively), the misalignment of LG and RG Dirac cones increases (decreases). Particularly, when $V_b$ increases negatively, the two Dirac cones are gradually aligned until $V_b = -0.5$ V (Fig. 2f, third left panel) and become misaligned again at $V_b < -0.5$ V (Fig. 2f, first and second left panels). The corresponding $T(E; V_b, V_g)$ spectra then show a strong resonant tunneling behavior at $V_b = -0.5$ V (Fig. 2f, third right panel) and a subsequent order of magnitude decrease at $V_b = -0.7$ V (Fig. 2f, second right panel), leading to a nonlinear $J - V_b$ characteristic with a sharp NDR peak (Fig. 2d, red curve). The peak-to-valley ratio (PVR) is about 2, which is in good agreement with the experimentally obtained values (up to $\approx 4$)[10]. On the other and, in the $V_b > 0$ V regime, the offset between LG and RG Dirac cones further increases (Fig. 2f, fifth left panel). The widening of the bias window for a nearly fixed $T(E; V_b, V_g)$ then results in a monatomic increase of current density without an NDR signal (but with a bump that occurs when $\mu_L$ crosses the transmission peak; Fig. 2f, fifth right panel), producing the asymmetric NDR $J - V_b$ characteristics shown in Fig. 2d.

## 2.3 Symmetric NDR $J$-$V_b$ curves from hBN point defects hybridized with graphene

It should be noted that the $J - V_b$ curves calculated above exhibit asymmetric features such that NDR peaks appear at positive (negative) $V_b$ under positive (negative) $V_g$. However, the symmetric NDR $J - V_b$ curve which has also been observed in experiment[10], cannot be explained in terms of the gating effect. We now incorporate atomic defects which inevitably occur in typical experimental conditions[43] and show that it provides a hitherto undiscussed mechanism that produces symmetric NDR $J - V_b$ characteristics. Here, we will set $V_g = 0$ V to make it clear that the identified NDR mechanisms are not related with gating effects. In addition, as a representative system, we will focus on the 3L hBN case with a defect in the middle hBN layer. As the point defect, we will adopt a substitutional carbon atom introduced on a nitrogen site ($C_N$), which is known as the dominant atomic defect type for hBN[30]. Then, as shown in Fig. 3a, we indeed find in the calculated $J - V_b$ characteristics NDR features symmetric with respect to $V_b$. Compared with the $V_g$-induced NDR $J - V_b$ curves shown in Fig. 2d, we find that the NDR peak appears in a lower $V_b$ regime (0.15 V) and the PVR is also smaller (1.42), whose details are overall consistent with experimental trends[10].

In addition to the NDR $J - V_b$ characteristic symmetric with respect to $V_b$, we also point out that as shown in Fig. 3b the $C_N$ defect induces a nonlinear voltage drop $\Delta \bar{v}_H$ profile that nontrivially varies with increasing $V_b$. Generally, we first note that, unlike in the pristine hBN case (Figs. 2b, e), the incorporation of $C_N$ defect results in nonuniformly distributed $\Delta \bar{v}_H$ profiles. It can be observed that at $V_b = 0.05$ V $\Delta \bar{v}_H$ drops more abruptly on the left graphene-hBN interface side (Fig. 3b, left top panels). However, as $V_b$ increases to 0.4 V, the region where $\Delta \bar{v}_H$ more abruptly drops moves to the lower electrochemical-potential right hBN-graphene interface side (Fig. 3b, right panel).

To fully understand the different electrostatic potential drop profiles at low and high $V_b$ regimes and its correlation with the NDR quantum transport characteristics, we analyzed as shown in Fig. 3c the electronic bands projected onto LG, hBN, and RG (left, center, and right panels, respectively). At zero-bias equilibrium state, as detailed in Supplementary Fig. 7b[44,45], the strong hybridization between $C_N$ and LG/RG states across one pristine hBN layer allows the charge transfer from LG/RG to $C_N$, making both LG and RG electrodes initially p-doped. Next, applying a small bias $V_b = 0.05$ V, we find that in-gap $C_N$ defect levels are pinned at the lower electrochemical potential $\mu_R$ (Fig. 3c left panel). The energetic proximity of $C_N$ defect states to $\mu_R$ then translates into their stronger coupling to RG states than to LG counterparts, explaining the more abrupt voltage drop at the LG-hBN interface side[28]. However, with increasing $V_b$, the defect state is pulled upward to the middle of the bias window at $V_b = 0.15$ V (Fig. 3c, center panel) and eventually to right below $\mu_L$ at $V_b = 0.4$ V (Fig. 3c, right panel). Namely, we find that with increasing $V_b$ electrons tunnel from LG across the left hBN layer into in-gap $C_N$ states located in the middle hBN layer. Upon the continued electron filling, energetically upshifting $C_N$ defect levels eventually approach $\mu_L$ and eventually couple more strongly to LG states than RG states, explaining the more abrupt voltage drop at the hBN-RG interface side at $V_b = 0.4$ V (Fig. 3b, right panel).

We now explain how the self-consistent charging of in-gap $C_N$ states induces the NDR $J - V_b$ feature shown in Fig. 3a. In Fig. 3d, the projected density of states (DOS) and transmission spectra corresponding to the band structures of Fig. 3c are presented (left and right panels, respectively). We first note that the $C_N$ defect-originated sharp DOS and $T(E; V_b)$ peaks are initially located at $\mu_R$ at $V_b = 0.05$ V (Fig. 3d, left panel), and with increasing $V_b$ energetically upshift toward $\mu_L$ (Fig. 3b, center and right panels). At this point, we remind that the $V_b$-induced charging of $C_N$ defect states indicates the strong hybridization between in-gap $C_N$ and graphene states as shown from the initial p-doping of LG/RG in equilibrium (Supplementary Fig. 7b), and the strength of $T(E; V_b)$ peaks quantifies the degree of quantum hybridizations. The energetic movement of the DOS and $T(E; V_b)$ peaks from the $\mu_R$ boundary into the middle of the bias window increases the current density until $V_b = 0.15$ V (NDR peak) (Fig. 3d, first and second right panels, respectively; see insets for zoom-in views). However, as $C_N$-originated in-gap defect levels continue to move upward with increasing $V_b$ and approach $\mu_L$ around $V_b = 0.4$ V (Fig. 3d, third left panel), it crosses the RG Dirac point where RG states are deficient and accordingly the transmission peak is reduced both in terms of height and width (NDR valley) (Fig. 3d, third right panel). Quantitatively, between the NDR peak ($V_b = 0.05$ V) and valley ($V_b = 0.4$ V), the $T(E; V_b)$ peak height decreases from 0.215 to 0.152 and the full width at half maximum ($> 10^{-3}$) decreases from 0.03 eV to 0.015 eV (insets of Fig. 3d second and third right panels).

We emphasize that the above-described non-equilibrium device electronic structures and quantum transport properties are significantly modified when the location of $C_N$ defect is different, signifying the importance of atomic details in the operation of 2D vdW electronic devices. Specifically, when a $C_N$ defect is introduced into the left interfacial hBN layer (rather than the middle hBN layer as discussed in Fig. 3), we find that currents decrease by an order of magnitude, $J - V_b$ characteristics become asymmetric, and importantly symmetric NDR features disappear (Fig. 4a, see also Supplementary Fig. 8). Analyses of non-equilibrium junction electronic structures show that in this case in-gap $C_N$ defect levels are strongly hybridized with LG states and pinned below the electrochemical potential $\mu_L$ (Figs. 4c-d), resulting in voltage drop profiles in which voltage always drops dominantly at the hBN-RG side (Fig. 4b). Bias-independent voltage drop profiles then nullify the above-described mechanism of defect-induced symmetric NDR $J-V_b$ characteristics, explaining the disappearance of NDR peaks. Moreover, as in the middle-hBN-layer $C_N$ case (see Supplementary Fig. 7b), charge transfers to the $C_N$ located in the left hBN layer induce at the equilibrium condition ($V_b = 0$ V) the strong and weak p-type doping of LG and RG electrodes, respectively (Supplementary Fig. 7c). Accordingly, unlike in the positively increasing $V_b$ case, the negatively increasing $V_b$ can align the single rings of $\vec{k}$ points in the LG and RG Dirac cones within the bias window and produce larger currents (see Supplementary Fig. 8).

*2.4 NDR characteristics from gating vs quantum-hybridization*

We close by providing mechanistic insights into the NDR characteristics originating from the gating-dependent tunneling and the quantum hybridization, whose differences are schematically summarized in Figs. 5a-b, respectively. The quantum-hybridization NDR is a general device concept involving the bias-induced breaking of initially delocalized quantum-hybridized states, and we previously predicted that a NDR with PVR > 10 can be achieved using atomically thin vdW nanowires with near-Ohmic homojunction contacts[46]. To reinforce the point defect-mediated quantum-hybridization NDR mechanism in graphene-based 2D vdW junctions, we visualize in Figs. 5c-d the NDR peak (left panels) and valley (right panels) local DOS around transmission peaks obtained from the pristine and $C_N$ defect-hBN cases, respectively. It can be noted that in the pristine hBN case (Fig. 5c) the in-phase (out-of-phase) local DOS are mainly located at the LG and RG electrode regions at the NDR peak (valley), reaffirming the tunneling mechanism. On the other hand, in the $C_N$ defect case (Fig. 5d), the local DOS have significant weights within the defective hBN region. Furthermore, the NDR peak (valley) is generated from the high (low) level of wavefunction delocalization, as evidenced by the presence (absence) of local DOS weight on the RG region at positive $V_b$ regimes.

## 3. CONCLUSION

To summarize, based on the recently developed MS-DFT formalism, we established an *ab initio* approach that enables the faithful modeling and simulation of 2D vdW FETs under finite-voltage operating conditions. The extended MS-DFT formalism then allowed in-depth atomistic analyses of two different NDR mechanisms that can emerge from graphene-based vertical 2D vdW FETs, for which theoretical studies have been so far mostly carried out at the semiclassical level employing the Bardeen transfer Hamiltonian formalism. Our explicit non-equilibrium first-principles calculations first quantified the mismatches between two graphene Dirac cones under finite gate- and/or bias-voltages, which are affected by the graphene quantum capacitance and produces asymmetric NDR $J$-$V_b$ characteristics. While this standard NDR mechanism concerns graphene electrodes, MS-DFT calculations revealed that atomic defects present within the hBN channel can also provide a hitherto undiscussed NDR mechanism. Specifically, we observed that for the $C_N$ defect placed in the middle hBN layer in-gap defect levels are self-consistently charged and upshifted with increasing $V_b$. Corresponding to bias-dependent changes of voltage drop profiles, the upshift of in-gap $C_N$ levels then reduced (enhanced) the coupling between $C_N$ and lower (higher) electrochemical-potential graphene states and produced symmetric NDR $J$-$V_b$ curves. On the other hand, with $C_N$ defects placed at outer interfacial hBN layers, in-gap $C_N$ levels were pinned to the adjacent graphene states or voltage drop profiles became bias-independent. These electronic structure characteristics then led to the disappearance of NDR peaks, decrease of currents by an order of magnitude, and asymmetric $J$-$V_b$ curves, signifying the critical importance of atomic-level details in the operation of 2D vdW devices. Our findings fill in the missing link in the present understanding of finite gate- and/or bias-voltage operations of vertical 2D vdW FETs, and also point towards future directions of computer-aided design of 2D nanodevices.

## ACKNOWLEDGEMENTS

This work was supported by the Samsung Research Funding & Incubation Center of Samsung Electronics (No. SRFC-TA2003-01). Computational resources were provided by KISTI Supercomputing Center (KSC-2018-C2-0032).

## AUTHOR CONTRIBUTIONS

Y.-H.K. developed the theoretical framework and oversaw the project. T.H.K., J.L., and R.G.L. implemented the method and carried out calculations. All authors analyzed the computational results, and Y.-H.K. wrote the manuscript with significant input from T.H.K and J.L.

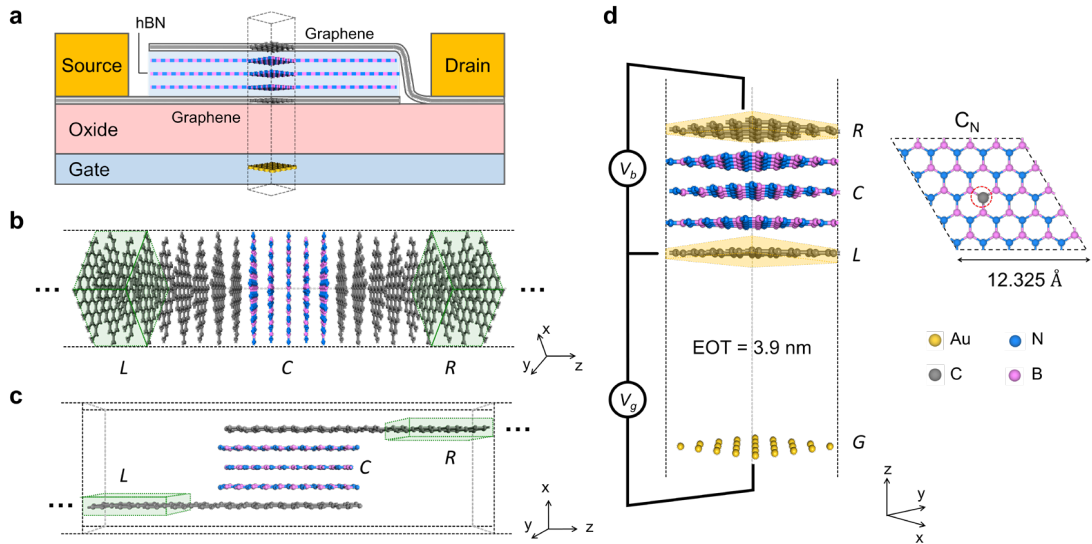

Fig. 1 Difficulties in *ab inito* modeling of graphene-based vertical FETs. **a** Schematic of the representative bottom gate graphene vertical FET device structure. **b** Graphite/hBN/graphite and **c** graphene/hBN nanoribbon/graphene junction models employed for DFT-NEGF calculations that require semi-infinite electrodes. Green shaded boxes indicate the electrode regions, for which within DFT-NEGF their matrix elements are replaced by those from separate bulk calculations. **d** The Au/vacuum/graphene/hBN/graphene vertical FET model employed in this work for MS-DFT calculations, which directly corresponds to the dashed box region in (**a**). Here, *L*, *C*, *R*, and *G* indicate the left electrode, channel, right electrode, and bottom gate electrode, respectively. Orange shaded boxes represent the electrode regions in MS-DFT calculations. A vacuum space between *L* and *G* was set to 10 Å, which corresponds to EOT of 3.9 nm.

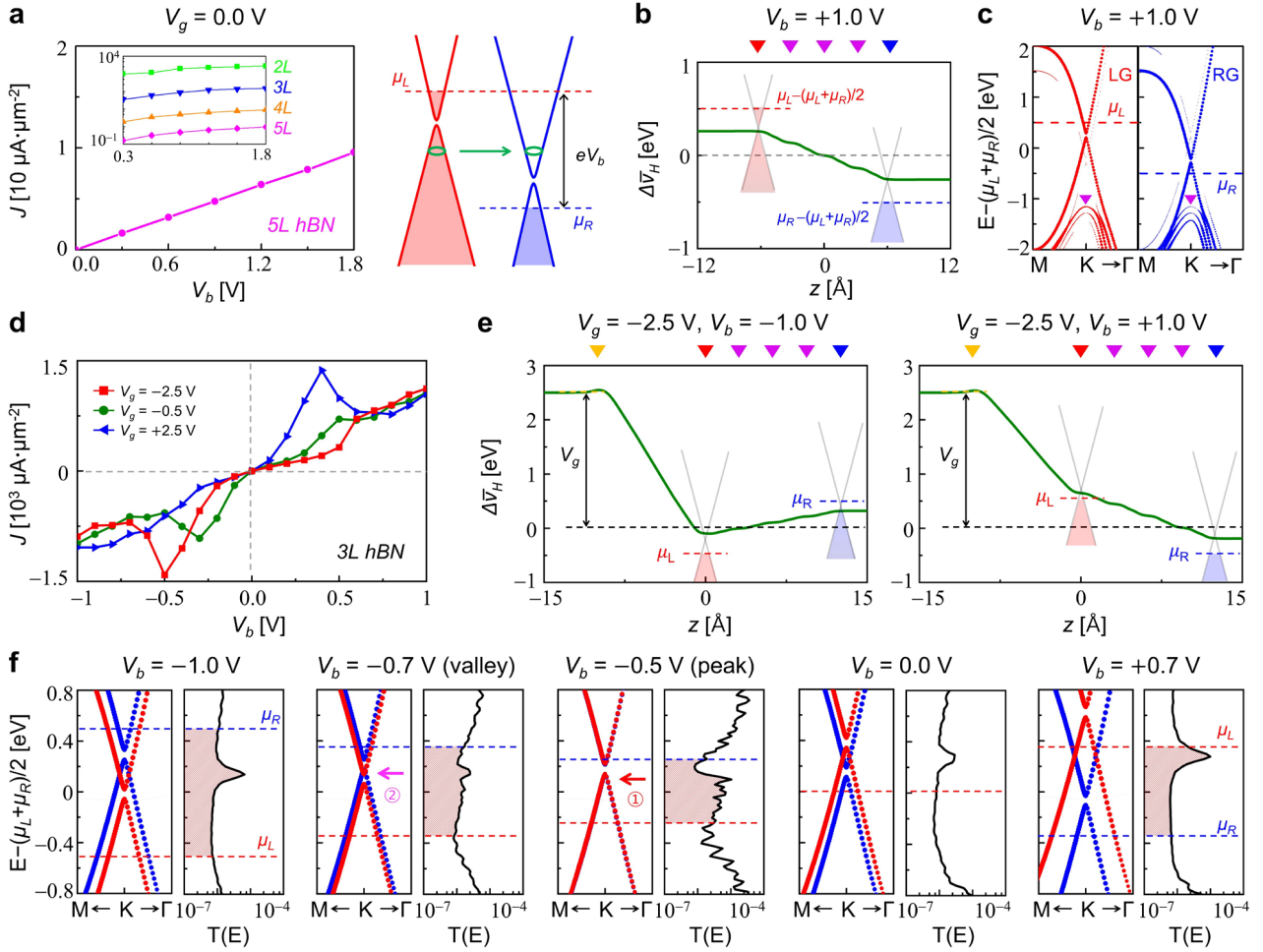

**Fig. 2 Gating-induced asymmetric NDR characteristics. a** The $V_g = 0$ V current density-bias voltage ($J - V_b$) characteristic of the left graphene (LG)/5L hBN/right graphene (RG) junction (left panel) and a schematic of the tunneling-based electron transport mechanism (right panel). Inset: The $J - V_b$ curves for the hBN 2—5L cases are shown on a logarithmic scale. **b** The $V_g = 0$ V $\Delta \bar{v}_H$ (upper) and $\Delta \bar{\rho}$ (lower) distributions of the LG/3L hBN/RG junction at $V_b$ = 1.0 V. Red, purple, and blue down triangles indicate the positions of LG, hBN, and RG layers, respectively. **c** The $V_g = 0$ V projected bands of the LG/3L hBN/RG junction at $V_b$ = 1.0 V. The circle size is in proportion to the weight of wavefunctions projected to the LG (red) and RG (blue) electrodes. Purple down triangles indicate the projections nearby hBN bands. **d** Gate-dependent $J - V_b$ curves of the LG/3L hBN/RG junction at $V_g = +2.5$ V (blue), $V_g = -0.5$ V (green), and $V_g = -2.5$ V (red). **e** The $V_g = -2.5$ V $\Delta \bar{v}_H$ profiles of the LG/3L hBN/RG junction at $V_b = -1.0$ V (left) and $V_b = +1.0$ V (right). Yellow, red, purple, and blue down triangles indicate the locations of the Au, LG, hBN, and RG layers, respectively. **f** The $V_g = -2.5$ V projected bands (left

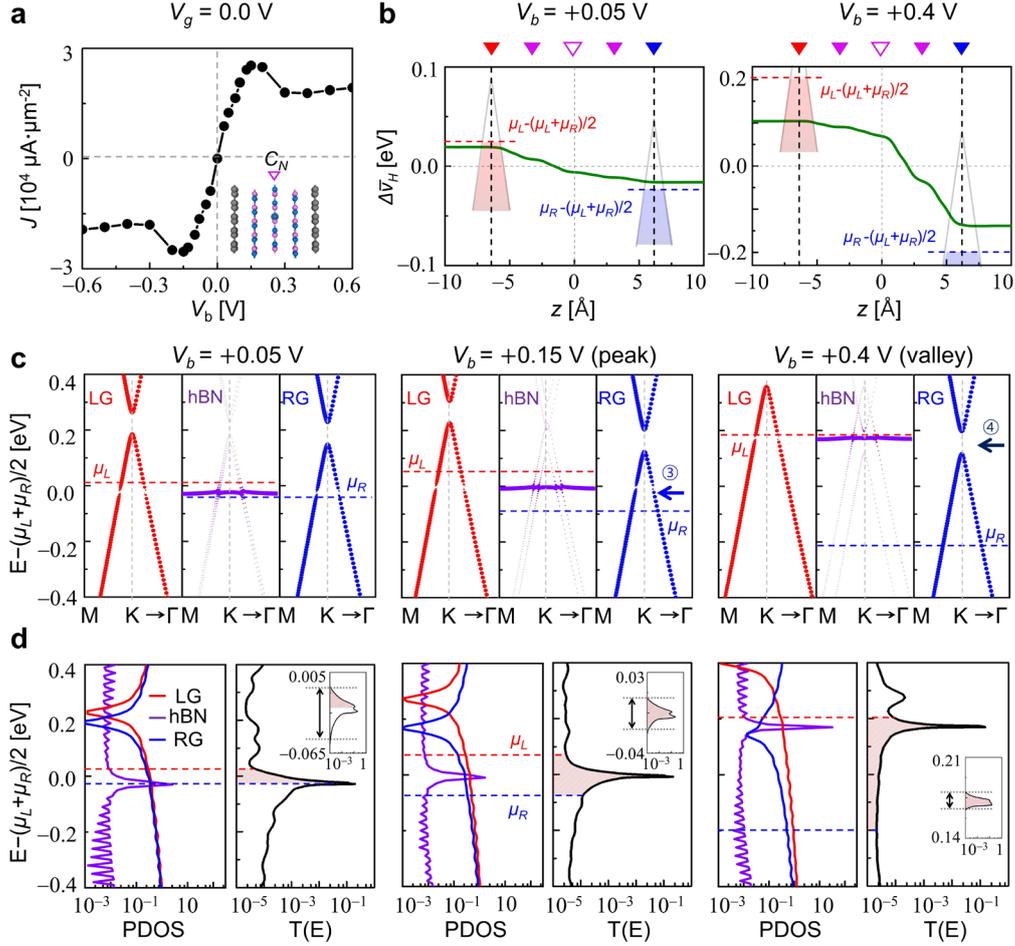

**Fig. 3 Symmetric NDR characteristics resulting from hBN defect levels. a** The zero-$V_g$ $J - V_b$ characteristic of LG/3L hBN/RG junction with a $C_N$ defect placed at the central hBN layer (inset). **b** The $\Delta \bar{v}_H$ distributions of the junction at $V_b =$ 0.05 V (left) and 0.4 V (right). Red, purple filled, purple empty, and blue down triangles indicate the positions of LG, pristine hBN, defective hBN, and RG layers, respectively. **c** The junction band structures projected onto LG (left panels), hBN (central panels), and RG (right panels) are shown for $V_b =$ 0.05 V, 0.15 V (NDR peak), and 0.4 V (NDR valley). The sizes of circles quantify the strength of orbital contributions. **d** The PDOS (left panels) and corresponding $T(E; V_b)$ spectra (rigth panels) at $V_b =$ 0.05 V, 0.15 V, and 0.4 V. Red, purple, and blue lines correspond to the LG, hBN, and RG states, respectively. Insets in (**d**): Zoomed-in views around the $T(E; V_b)$ peaks.

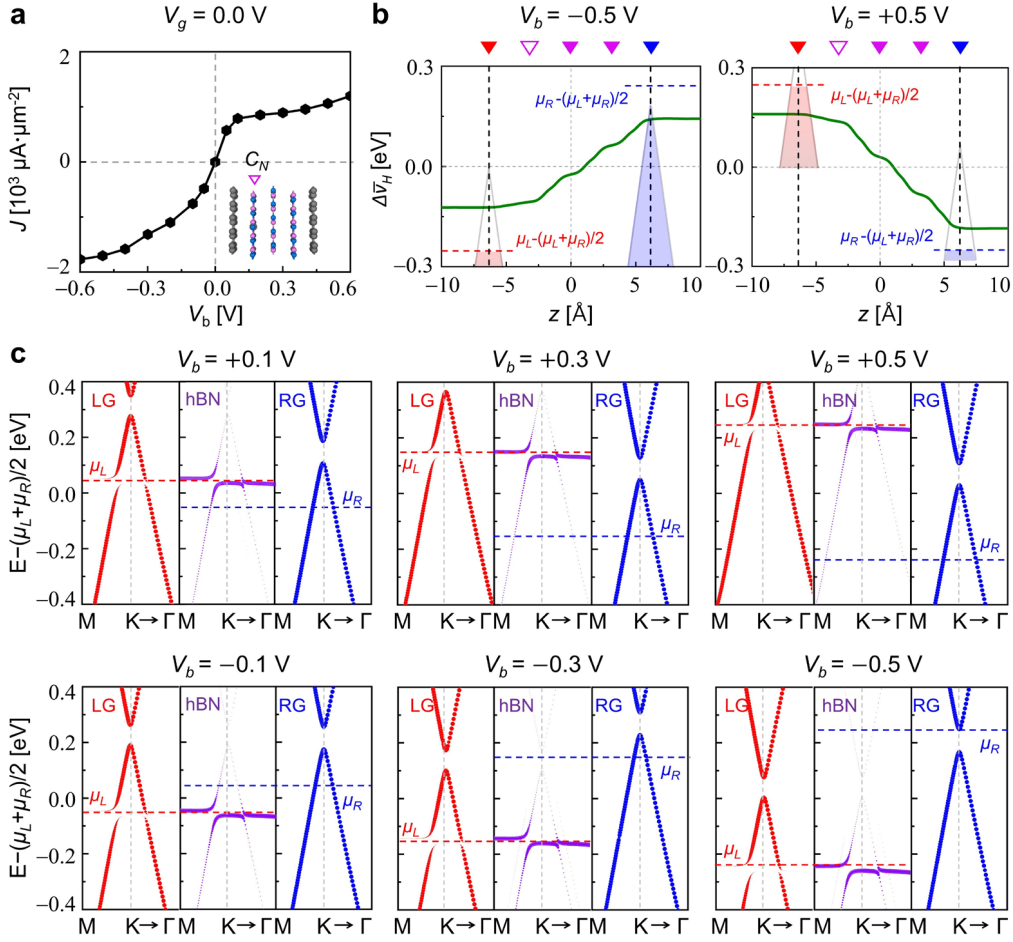

**Fig. 4 Critical effects of the defect location. a** The zero-$V_g$ $J - V_b$ characteristic of LG/3L hBN/RG junction with a $C_N$ defect placed at the left-most hBN layer (inset). **b** The $\Delta \bar{v}_H$ distributions at $V_b = +0.5$ V (left) and $-0.5$ V (right). Red, purple filled, purple empty, and blue down triangles indicate the positions of LG, pristine hBN, defective hBN, and RG layers, respectively. **c** The junction band structures projected onto LG (left panels), hBN (central panels), and RG (right panels) are shown for $V_b$ = +0.1 V, +0.3 V, +0.5 V (upper panels) and $V_b = -0.1$ V, $-0.3$ V, $-0.5$ V (lower panels). The sizes of circles quantify the strength of orbital contributions.

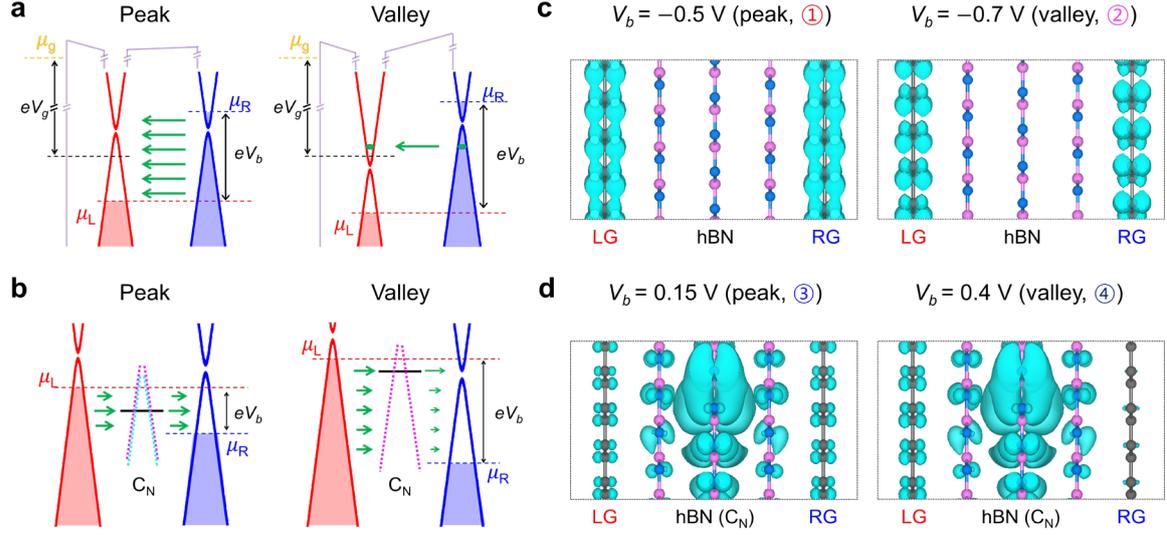

**Fig. 5 Schematics of the NDR mechanisms based on the gating and quantum hybridization. a** Negatively gated ($V_g = -2.5$ V) pristine 3L hBN case at the NDR peak ($V_b = -0.5$ V, left panel) and NDR valley ($V_b = -0.7$ V, right panel). **b** Defective 3L hBN case (central $C_N$ defect) at the NDR peak ($V_b = 0.15$ V, left panel) and NDR valley ($V_b = 0.4$ V, right panel). The red and blue solid lines indicate the LG and RG Dirac cones, respectively, and green arrows represent the electron flow. The chemical potentials of LG ($\mu_L$) and RG ($\mu_R$) that define the bias window with the voltage $V_b$ are indicated by the red and blue horizontal dotted lines. In (**b**), pink (cyan) dashed lines represent the hybridization of LG (RG) states with the $C_N$ defect level (black solid line). The local DOS for **c** the pristine 3L hBN case at $V_b = -0.5$ V (left panel) and $V_b = -0.7$ V (right panel), and for **d** the defective 3L hBN case at $V_b = 0.15$ V (left panel) and $V_b = 0.4$ V (right panel). The local DOS were obtained from the energy levels corresponding to transmission peaks (marked as ①, ②) in the third and second panels of Fig. 2**f**, respectively, for the pristine 3L hBN case, and as ③, ④ in the second and third panels of Fig. 3**c**, respectively, for the defective 3L hBN case). Energy windows of $[-0.05 \text{ eV}, +0.05 \text{ eV}]$ around the transmission peaks and the isosurface value of $1 \times 10^{-5}$ ($1 \times 10^{-4}$) states·Å$^{-3}$ were adopted for the pristine (defective) 3L hBN case.